\documentclass{article}
\usepackage{spconf,amsmath,graphicx,hyperref}
\usepackage{booktabs}
\usepackage{multirow}
\usepackage[table]{xcolor}
\usepackage{microtype}
\usepackage{spconf,amsmath,graphicx,hyperref}
\usepackage{xurl}
\definecolor{bestblue}{RGB}{0,70,180}
\definecolor{worstred}{RGB}{200,30,30}
\newcommand{\best}[1]{\textcolor{bestblue}{\textbf{#1}}}
\newcommand{\worst}[1]{\textcolor{worstred}{#1}}
\newcommand{\sw}[2]{{\setlength{\fboxsep}{1.2pt}\colorbox{#1}{#2}}}
\newcommand{\rhoal}{\ensuremath{\rho_{\mathrm{align}}}}
\newcommand{\effdim}{\ensuremath{d_{\mathrm{eff}}}}
\title{Rethinking Automated Voice Similarity by Shifting from EER to Embedding Geometry}
\name{Szu-Chi Chen$^{1}$, Jia-Kai Dong$^{1}$, Yi-Cheng Lin$^{1}$, Sung-Feng Huang$^{2}$, Hung-yi Lee$^{1,3}$}
\address{
    $^1$ National Taiwan University, Taipei, Taiwan \quad $^2$ NVIDIA, Taiwan \\
    $^3$ Artificial Intelligence Center of Research Excellence (NTU AI-CoRE), NTU, Taiwan \\
}
\begin{document}
\ninept
\maketitle
\begin{abstract}
Speaker verification (SV) models are commonly assumed to better capture nuances among speaker characteristics as verification accuracy improves, leading to their widespread use as automated proxies for human voice similarity in speech generation tasks. However, by establishing a human perceptual alignment metric and conducting systematic analysis, we demonstrate that perceptual alignment is governed far more by how a model is trained (its learning objective) than by how well it performs (EER). Notably, standard margin-based classification losses (e.g., AAM-Softmax) yield substantially lower perceptual alignment than prototypical metric losses, while EER itself fails to track human judgment—directly challenging the community's implicit assumption. We trace this divergence to embedding geometry, where a model's effective dimensionality ($d_{\mathrm{eff}}$) tracks perceptual alignment with a $-0.95$ rank correlation, revealing that the dimensional spread favored by classification losses fundamentally clashes with the low-dimensional nature of human voice perception. Imposing a dimensionality bottleneck compresses $d_{\mathrm{eff}}$ and raises perceptual alignment ($\rho_{\mathrm{align}}$) from $0.08$ to $0.74$, establishing a principled geometric criterion for evaluating voice similarity.
\end{abstract}
\begin{keywords}
Speaker embeddings, perceptual similarity, speaker verification, effective dimensionality, evaluation metrics
\end{keywords}
\section{Introduction}
\label{sec:intro}

Currently, in text-to-speech (TTS) and voice conversion (VC) systems, whether a model can accurately reproduce the reference speaker's timbre is one of the key target functionality. 
Because human scoring are scarce and expensive \cite{deja2022similarity}, practical evaluations typically rely on the speaker embedding cosine similarity as the metric to measure the similarity between synthesized speech and the reference speaker \cite{valle1,valle2,seedtts,cosyvoice}. 
To select the best model to extract the speaker embeddings, speaker verification (SV) is mainly chosen as the evaluation task.
Note that human perception of voice similarity is a continuous judgment that distinguishes between ``strongly similar'' and ``slightly similar.''
In contrast, the SV task simply distinguishes whether two samples belong to the same identity or not. 
Therefore, how the speaker representations' SV performance can reflect fine-grained human perceptual similarity remains largely unexplored.

In this study, we investigate the relationship between verification performance and human perceptual alignment under common speaker embedding model training setups that rely exclusively on identity labels without perceptual supervision. Specifically, we address two research questions:
\begin{itemize}
    \item \textbf{RQ1:} Does a better speaker verifier, as measured by EER, also achieve higher human perceptual alignment?
    \item \textbf{RQ2:} What property of a trained speaker embedding model tracks its perceptual alignment?
\end{itemize}

To address these questions, this study systematically evaluates the relationship between verification performance and human perceptual alignment across varying model conditions and training objectives. To further understand what drives perceptual alignment, we investigate the geometric properties of the learned embedding spaces to find metrics that track this alignment. Finally, we explicitly manipulate the embedding geometry to verify its direct impact on human perception.
The main contributions of this work can be summarized as follows:
\begin{itemize}
    \item \textbf{EER is not a reliable proxy for human perceptual alignment.} 
    The Spearman correlation between EER and perceptual alignment is only $+0.07$ (Sec.~\ref{ssec:study_a}).
    
    \item \textbf{Training objective determines perceptual alignment.} 
    In every model condition, the best and the worst objective differ by more than a factor of three in human perceptual alignment, without an EER cost (Sec.~\ref{ssec:study_a}).
    
    \item \textbf{Effective dimensionality tracks perceptual alignment.} 
    We test \effdim{}, how many directions a model spreads its speakers over. Its Spearman correlation with perceptual alignment is $-0.95$ (Sec.~\ref{ssec:study_e}).

    \item \textbf{Constraining the embedding dimension raises alignment.}
    On ECAPA-TDNN, narrowing the embedding dimension of AM-Softmax, the worst-aligned objective in the experiment, lowers \effdim{} and raises perceptual alignment correlation from $0.08$ to $0.74$ (Sec.~\ref{ssec:study_g}).
\end{itemize}

\begin{figure}[htb]
\centering
\centerline{\includegraphics[width=\linewidth]{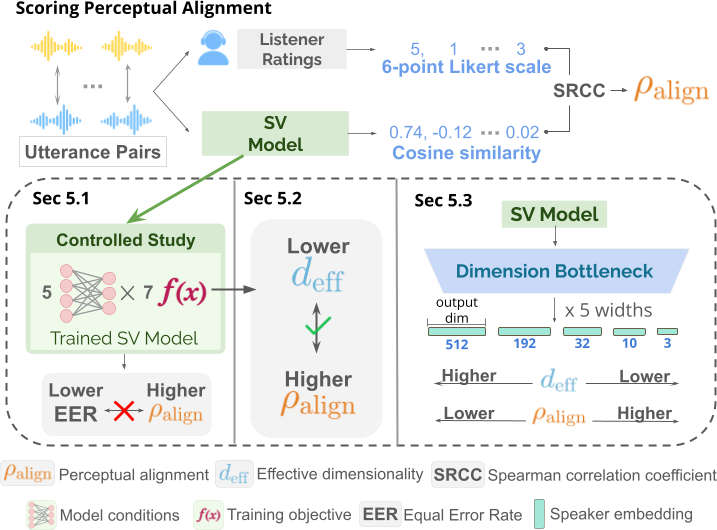}}
\caption{Overview of our study. Top: measurement of perceptual alignment; bottom: controlled experiments, effective dimensionality analysis, and embedding dimension bottlenecks.}
\label{fig:workflow}
\end{figure}

\section{Related Work}
\label{sec:related_work}

\subsection{Speaker Embedding Models, Speaker Verification and Human Perceptual Similarity}

Whether speaker embeddings can estimate human listening perception is not a new question. Earlier studies compared model scores with listener ratings to automatically select acoustically similar speakers~\cite{gerlach2020similarity,liu2024comparison}. 
Afterward, as speaker embedding models became standard evaluation tools in TTS and VC, the guiding question shifted from whether these models can substitute human listeners to which model serves as the best proxy ~\cite{das2020predictions}.
Since the SV task routinely evaluates how well a speaker embedding model captures the differences between unseen speakers~\cite{chung2020defence}, the community often takes good verification accuracy as a sign of well-generalized perceptual capability and therefore defaults to the best-performing SV model (with the lowest EER)~\cite{valle1} to estimate human perception, which is a foundational premise that remains empirically unexamined.

\subsection{Speaker Embedding Training Objectives}
Early speaker embeddings were not explicitly optimized for embedding similarity: i-vectors~\cite{i-vector} were learned with an unsupervised generative criterion, and x-vectors~\cite{snyder2018xvector} emerged as a by-product of softmax speaker classification~\cite{chung2020defence}.
In contrast, modern objectives are designed to optimize embedding similarity, and fall into two families.
The classification family, including AM-Softmax~\cite{wang2018amsoftmax} and AAM-Softmax~\cite{deng2019arcface}, normalizes embeddings and imposes a margin on their cosine similarity to learnable class centers;
while the metric learning family, including GE2E~\cite{wan2018ge2e}, Prototypical~\cite{snell2017proto}, Angular Prototypical~\cite{chung2020defence} and supervised contrastive loss~\cite{khosla2020supcon}, instead compares embeddings with each other directly or with class centroids computed from the batch.
Although they differ in what each embedding is compared against, both families shape the similarity structure of the embedding space, making cosine similarity the standard scoring function for estimating speaker similarity.
While \cite{chung2020defence} comprehensively benchmarked most of these objectives on verification accuracy, how they affect alignment with human perceptual similarity remains unexplored.

\subsection{Representation Geometry and Dimensionality}

Cosine similarity measures the angle between embeddings within the subspace that the representations span. The effective dimensionality of these representations can be much lower than their nominal embedding dimension, with different training objectives retaining different degrees of directional variation \cite{jing2022dimcollapse, roth2020revisiting}. This geometric perspective is conceptually consistent with perceptual studies showing that human judgments of voice similarity can be captured by low-dimensional spaces \cite{baumann2010perceptual}. Representation dimensionality thus emerges as a potential geometric link between speaker embeddings and human perceptual similarity. While prior work has documented how objectives reshape representation dimensionality, its direct connection to human perceptual alignment remains insufficiently understood, a gap this work explicitly addresses.

\section{Method}
\label{sec}

\subsection{Scoring Perceptual Alignment}
\label{ssec:perceptual_alignment}
As illustrated in Fig.~\ref{fig:workflow}, we measure \emph{perceptual alignment} (\rhoal) as the Spearman rank correlation coefficient (SRCC) between speaker embedding cosine similarities and mean listener ratings across all different-speaker pairs in VoxSim.
For each utterance pair $(i,j)$, let $s_{ij}$ denote the cosine similarity between their speaker embeddings and $r_{ij}$ denote the corresponding mean listener rating. We define
\begin{equation}
\rho_{\mathrm{align}}
=
\operatorname{SRCC}
\left(
\{s_{ij}\}_{(i,j)\in\mathcal{P}_{\mathrm{diff}}},
\{r_{ij}\}_{(i,j)\in\mathcal{P}_{\mathrm{diff}}}
\right),
\end{equation}
where $\mathcal{P}_{\mathrm{diff}}$ contains the different-speaker pairs in VoxSim.

We restrict evaluation to different-speaker pairs to measure \emph{graded} perceptual similarity rather than identity discrimination. Including same-speaker pairs introduces a strong binary identity signal that can inflate correlation without capturing similarity among different speakers.\footnote{Across all rated VoxSim pairs, an identity-only baseline assigning 1 to same-speaker and 0 to different-speaker pairs already achieves an SRCC of 0.715. Restricting evaluation to the 16,905 different-speaker pairs removes this shortcut and better separates embedding models.}

\subsection{Effective dimensionality}
\label{ssec:effdim}

To test whether representation dimensionality explains the differences in human perceptual alignment scores across training objectives, we require a metric that directly quantifies the intrinsic geometric dimensionality of the embedding space. We adopt effective dimensionality (\effdim) \cite{gao2017theory, delgiudice2021ed} to measure the extent of dimensional spread in the embedding space.

For each trained model, we L2-normalize the VoxSim utterance embeddings before and after averaging them by speaker into $1{,}251$ speaker centroids,\footnote{Using centroids suppresses utterance noise and isolates between-speaker geometry, ensuring the metric's scope strictly aligns with our cross-speaker perceptual evaluation.}  then center the centroids and compute the participation ratio (PR) of their PCA eigenvalue spectrum\footnote{Compared with alternative dimensionality metrics, the PR is less sensitive to long-tail noise eigenvalues and provides a more conservative estimate \cite{delgiudice2021ed}.} \cite{gao2017theory}:
\begin{equation}
  \effdim = \frac{\bigl(\sum_{i=1}^{K}\lambda_i\bigr)^{2}}
                 {\sum_{i=1}^{K}\lambda_i^{2}},
\label{eq:effdim}
\end{equation}
where $\lambda_1\!\ge\!\cdots\!\ge\!\lambda_K$ are the eigenvalues of the
$K\!\times\!K$ centroid covariance matrix ($K$ = embedding dimensionality).

Intuitively, \effdim{} is the equivalent number of equal-variance orthogonal
directions producing the same pattern of covariation \cite{delgiudice2021ed}:
larger means a more uniform speaker space. It behaves as a continuous counterpart of matrix rank --- the effective number of independent
directions over which a model spreads its speakers.

\begin{table*}[!t]
\caption{The controlled experimental matrix: EER (\%, VoxCeleb1-O), perceptual alignment \rhoal{}, and effective dimensionality \effdim{} (Eq.~\ref{eq:effdim}); mean over 3 seeds ($\pm$ sd for EER and \rhoal{}; \effdim{} seed sd $\le 4.5$). \best{Blue bold} and \worst{red} mark the highest and lowest \rhoal{} per model condition; \textbf{bold} marks the best EER.}
\label{tab:main}
\centering
\footnotesize
\setlength{\tabcolsep}{1.3pt}
\begin{tabular*}{\textwidth}{@{\extracolsep{\fill}}lccccccccccccccc@{}}
\toprule
 & \multicolumn{3}{c}{ECAPA-TDNN} & \multicolumn{3}{c}{Fast ResNet-34} & \multicolumn{3}{c}{ReDimNet-B2} & \multicolumn{3}{c}{WavLM-p1} & \multicolumn{3}{c}{WavLM-p2} \\
\cmidrule(lr){2-4}\cmidrule(lr){5-7}\cmidrule(lr){8-10}\cmidrule(lr){11-13}\cmidrule(lr){14-16}
loss & EER$\downarrow$ & \rhoal$\uparrow$ & \effdim & EER$\downarrow$ & \rhoal$\uparrow$ & \effdim & EER$\downarrow$ & \rhoal$\uparrow$ & \effdim & EER$\downarrow$ & \rhoal$\uparrow$ & \effdim & EER$\downarrow$ & \rhoal$\uparrow$ & \effdim \\
\midrule
\multicolumn{16}{l}{\emph{Metric: prototypical}} \\
Prototypical & 1.92{\tiny$\pm$.04} & .395{\tiny$\pm$.010} & 27.1 & 2.39{\tiny$\pm$.13} & .491{\tiny$\pm$.034} & 18.3 & 1.46{\tiny$\pm$.05} & .423{\tiny$\pm$.001} & 27.9 & 1.75{\tiny$\pm$.09} & \best{.462}{\tiny$\pm$.029} & 21.2 & 1.37{\tiny$\pm$.04} & \best{.428}{\tiny$\pm$.011} & 25.2 \\
Angular Proto. & \textbf{1.47}{\tiny$\pm$.04} & \best{.406}{\tiny$\pm$.007} & 33.0 & 2.35{\tiny$\pm$.12} & \best{.499}{\tiny$\pm$.012} & 21.2 & \textbf{1.15}{\tiny$\pm$.01} & \best{.425}{\tiny$\pm$.011} & 31.0 & 1.42{\tiny$\pm$.06} & .455{\tiny$\pm$.024} & 26.1 & \textbf{1.09}{\tiny$\pm$.10} & .401{\tiny$\pm$.003} & 31.2 \\
\midrule
\multicolumn{16}{l}{\emph{Classification}} \\
NSL & 2.10{\tiny$\pm$.03} & .207{\tiny$\pm$.009} & 59.1 & 2.82{\tiny$\pm$.14} & .249{\tiny$\pm$.016} & 45.9 & 1.82{\tiny$\pm$.11} & .171{\tiny$\pm$.006} & 67.4 & 1.83{\tiny$\pm$.10} & .298{\tiny$\pm$.012} & 44.2 & 2.31{\tiny$\pm$.09} & .215{\tiny$\pm$.015} & 54.9 \\
AAM-Softmax & \textbf{1.47}{\tiny$\pm$.12} & .111{\tiny$\pm$.013} & 74.5 & \textbf{2.18}{\tiny$\pm$.11} & .239{\tiny$\pm$.022} & 40.6 & 1.52{\tiny$\pm$.08} & .143{\tiny$\pm$.014} & 62.9 & 1.26{\tiny$\pm$.07} & .205{\tiny$\pm$.005} & 53.9 & 1.61{\tiny$\pm$.13} & .152{\tiny$\pm$.017} & 60.8 \\
AM-Softmax & 1.62{\tiny$\pm$.05} & \worst{.080}{\tiny$\pm$.009} & 98.3 & 2.19{\tiny$\pm$.11} & \worst{.158}{\tiny$\pm$.017} & 64.0 & 1.60{\tiny$\pm$.09} & \worst{.094}{\tiny$\pm$.009} & 89.7 & \textbf{1.25}{\tiny$\pm$.03} & \worst{.144}{\tiny$\pm$.012} & 71.7 & 1.40{\tiny$\pm$.03} & \worst{.102}{\tiny$\pm$.011} & 80.5 \\
\midrule
\multicolumn{16}{l}{\emph{Metric: other}} \\
GE2E & 1.59{\tiny$\pm$.08} & .282{\tiny$\pm$.026} & 41.1 & 2.35{\tiny$\pm$.11} & .346{\tiny$\pm$.009} & 28.5 & 1.52{\tiny$\pm$.31} & .331{\tiny$\pm$.018} & 36.0 & 1.36{\tiny$\pm$.12} & .341{\tiny$\pm$.012} & 31.8 & 1.11{\tiny$\pm$.09} & .280{\tiny$\pm$.003} & 37.2 \\
SupCon & 5.59{\tiny$\pm$.64} & .295{\tiny$\pm$.060} & 34.3 & 3.54{\tiny$\pm$.22} & .352{\tiny$\pm$.016} & 27.0 & 3.87{\tiny$\pm$.46} & .327{\tiny$\pm$.013} & 26.2 & 2.79{\tiny$\pm$.02} & .356{\tiny$\pm$.019} & 30.4 & 2.67{\tiny$\pm$.16} & .312{\tiny$\pm$.012} & 35.1 \\
\bottomrule
\end{tabular*}
\end{table*}

\section{Experimental setup}
\label{sec:setup}

\subsection{Models and training}
\label{ssec:conditions}

We vary the training objective inside five model conditions that differ in
architecture family, input representation, and embedding width. Three of them
take log-mel filterbank input and train for 80 epochs: ECAPA-TDNN
\cite{desplanques2020ecapa} and ReDimNet-B2 \cite{yakovlev2024redimnet} with
192-dimensional embeddings, and Fast ResNet-34 \cite{chung2020defence} with
512-dimensional embeddings. The other two build on a WavLM Base+ trunk, whose
layer outputs are combined by a learnable weighted sum and fed to an ECAPA-TDNN
head that produces a 192-dimensional embedding, adapting the two-stage recipe of
\cite{chen2022wavlm}. \emph{WavLM-p1} trains the head for 20 epochs on a frozen
trunk; \emph{WavLM-p2} then unfreezes the WavLM transformer encoder and trains for
5 more epochs. Both stages are reported as separate conditions.

All models train on VoxCeleb2-dev \cite{chung2018voxceleb2}, 5{,}994
speakers.
Two main settings are used in our experiments. In Sec.~\ref{ssec:study_a}, we vary the training objective inside each of the five model conditions, where every condition is trained with the seven objectives described next, three seeds each. In Sec.~\ref{ssec:study_g}, we then varies the embedding dimension: ECAPA-TDNN is chosen and re-trained at $n_{\mathrm{out}}\in\{512,32,10,3\}$ with four of those objectives, AAM-Softmax, AM-Softmax, Angular Prototypical, and Prototypical, three seeds each.
Code, configurations, and seeds are released.\footnote{\url{https://github.com/47zzz/voice-similarity-embedding-geometry}}

\subsection{Training objectives}
\label{ssec:objectives}

For our experiment in Sec.~\ref{ssec:study_a}, we incorporate seven training objectives, comprising three classification-based and four metric-learning losses. 

\textbf{Classification.} We consider normalized softmax loss (NSL) \cite{wang2018cosface}, AM-Softmax \cite{wang2018amsoftmax}, and AAM-Softmax \cite{deng2019arcface}. NSL serves as the baseline without a margin ($m{=}0$), whereas AM-Softmax and AAM-Softmax incorporate an explicit margin with $m{=}0.2$.

\textbf{Metric.} We include Prototypical loss and Angular Prototypical loss \cite{snell2017proto,chung2020defence}, GE2E \cite{wan2018ge2e}, and SupCon \cite{khosla2020supcon}. The two prototypical variants share the same formulation, which averages a speaker's other utterance embeddings into one reference and requires a held-out utterance to lie closest to the reference of its own speaker.

Batch sizes follow the batch-size study of \cite{chung2020defence}:
classification objectives use their best-performing fixed batch of 200
utterances, and metric objectives, which they found to benefit from larger
batches, use batches up to $N{=}400$ speakers $\times$ $M{=}2$ utterances.

\subsection{Evaluation data}
\label{ssec:Evaluation data}

Two sets serve for evaluation, and neither shares a speaker with the training corpus. The first is the cleaned VoxCeleb1-O trial list \cite{nagrani2017voxceleb,nagrani2020voxceleb}, the standard verification benchmark. The second is VoxSim \cite{ahn2024voxsim}: 46{,}348 utterances from 1{,}251 VoxCeleb1 speakers, on which 13 listeners rated pairs of recordings for how similar the two speakers sound, on an integer 1--6 scale with about two ratings per pair; we merge the ratings into listener means over 27{,}697 unique pairs, of which 16,905 are different-speaker and 10,792 same-speaker pairs. We report standard EER on VoxCeleb1-O using raw cosine similarity between full-recording embeddings.

\begin{table}[htb]
\caption{Spearman rank correlation of \rhoal{} with EER (\%, VoxCeleb1-O) and
with \effdim{}, each training condition one observation (three seeds averaged).
Parentheses: two-sided permutation-test $p$-values, exact for $n{=}7$ and Monte
Carlo ($2{\times}10^{6}$ permutations) for $n{=}35$.}
\label{tab:perarch}
\centering
\footnotesize
\setlength{\tabcolsep}{4pt}
\providecommand{\pv}[1]{{\scriptsize$(p#1)$}}
\begin{tabular}{lccc}
\toprule
Population & $n$ & Spearman$(\mathrm{EER},\rhoal)$ & Spearman$(\effdim,\rhoal)$ \\
\midrule
\multicolumn{4}{l}{\emph{Within model conditions}} \\
ECAPA-TDNN  & 7 & $+0.11$\,\pv{=.84}  & $-0.96$\,\pv{=.003} \\
Fast ResNet-34 & 7 & $+0.43$\,\pv{=.35}  & $-0.93$\,\pv{=.007} \\
ReDimNet-B2    & 7 & $-0.68$\,\pv{=.11}  & $-0.71$\,\pv{=.088} \\
WavLM-p1    & 7 & $+0.61$\,\pv{=.17}  & $-1.00$\,\pv{<.001} \\
WavLM-p2    & 7 & $-0.36$\,\pv{=.44}  & $-1.00$\,\pv{<.001} \\
\midrule
\multicolumn{4}{l}{\emph{Pooled across conditions}} \\
All conditions & 35 & $+0.07$\,\pv{=.71} & $-0.95$\,\pv{<.001} \\
\bottomrule
\end{tabular}
\end{table}

\section{Results and Analysis}
\label{sec:studies}

\subsection{Lower EER Does Not Imply Higher Perceptual Alignment}
\label{ssec:study_a}

\begin{figure}[!t]
\centering
\includegraphics[width=\columnwidth]{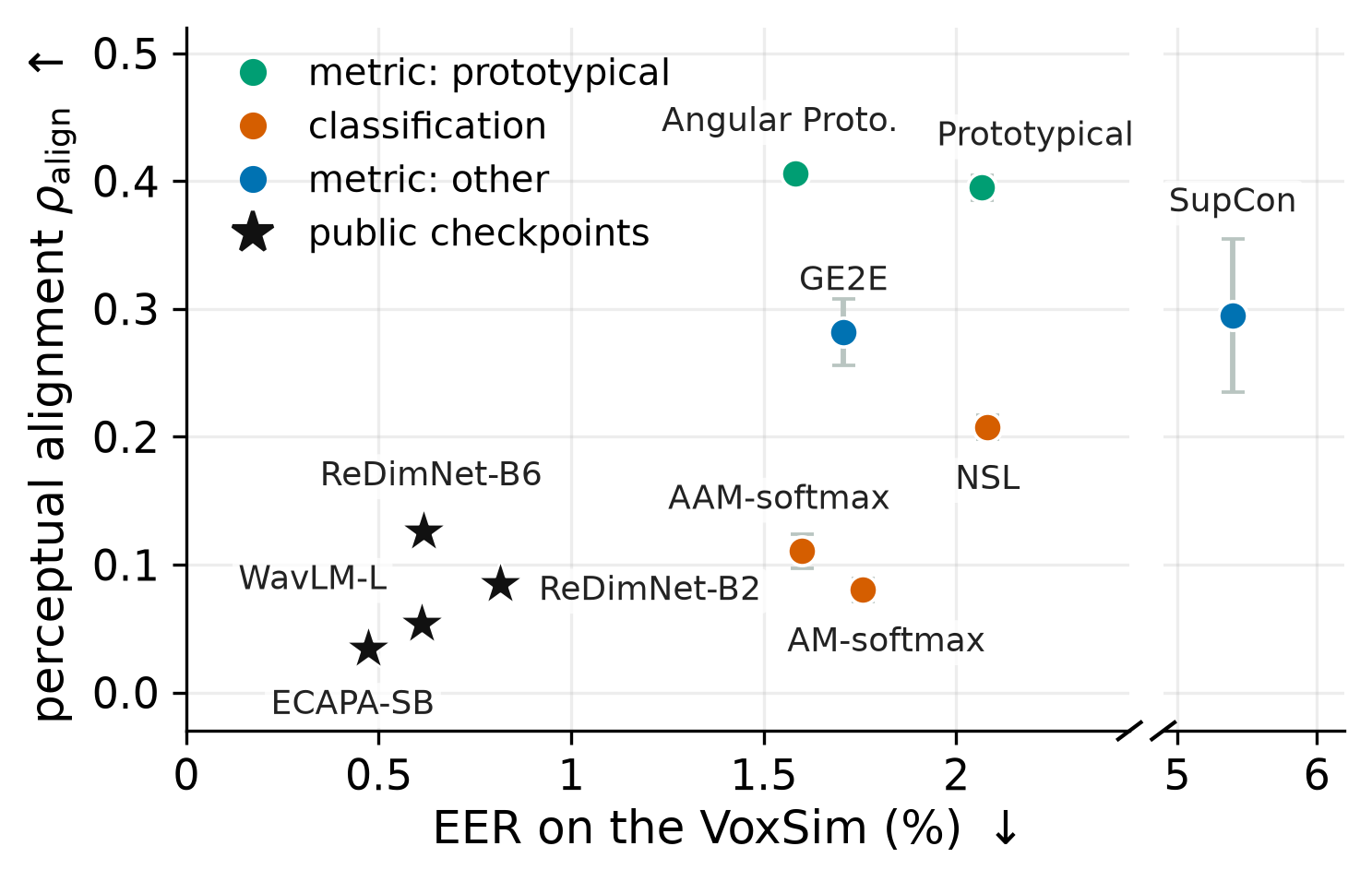}
\caption{Verification vs.\ perceptual alignment on ECAPA (circles; 3-seed mean $\pm$ sd) and the four public checkpoints (stars), with EER scored on the VoxSim pairs.}
\label{fig:dissociation}
\end{figure}

Table~\ref{tab:main} reports the main experimental matrix. Across the 35 training conditions, $\textit{Spearman}(\mathrm{EER},\rhoal)=+0.07$, and within single model conditions the sign is unstable, from $-0.68$ to $+0.61$, with no permutation test rejecting zero correlation (p-value $p\geq.11$; Table~\ref{tab:perarch}).
Fig.~\ref{fig:dissociation} plots both metrics on the VoxSim pairs across identical audio recordings. Focusing on the ECAPA model condition, we examine our trained models alongside four public checkpoints marked by stars~\cite{yakovlev2024redimnet,chen2022wavlm,desplanques2020ecapa,ravanelli2021speechbrain}. Overall, the plot demonstrates a clear dissociation between EER and perceptual alignment. Crucially, the public state-of-the-art checkpoints land in the lower-left region with low EER but poor perceptual alignment, contradicting the community expectation that better speaker verification leads to higher human perceptual alignment.

While EER is uninformative about human perception, Table~\ref{tab:main} shows that the training objective largely determines perceptual alignment. In every model condition, the best and worst objectives differ in \rhoal{} by over a factor of three, without incurring EER cost. Specifically, prototypical losses consistently achieve the highest \rhoal{} (0.395--0.499), whereas classification losses yield the lowest (0.080--0.298), with AM-Softmax universally last. For instance, under identical conditions on ECAPA, replacing AAM-Softmax with Angular Prototypical nearly quadruples \rhoal{} (from $0.111\pm0.013$ to $0.406\pm0.007$) while maintaining a tied-best EER of 1.47\%. Substantial gains in perceptual alignment thus demand no compromise in verification performance.

\begin{figure}[htb]
\centering
\centerline{\includegraphics[width=0.95\linewidth]{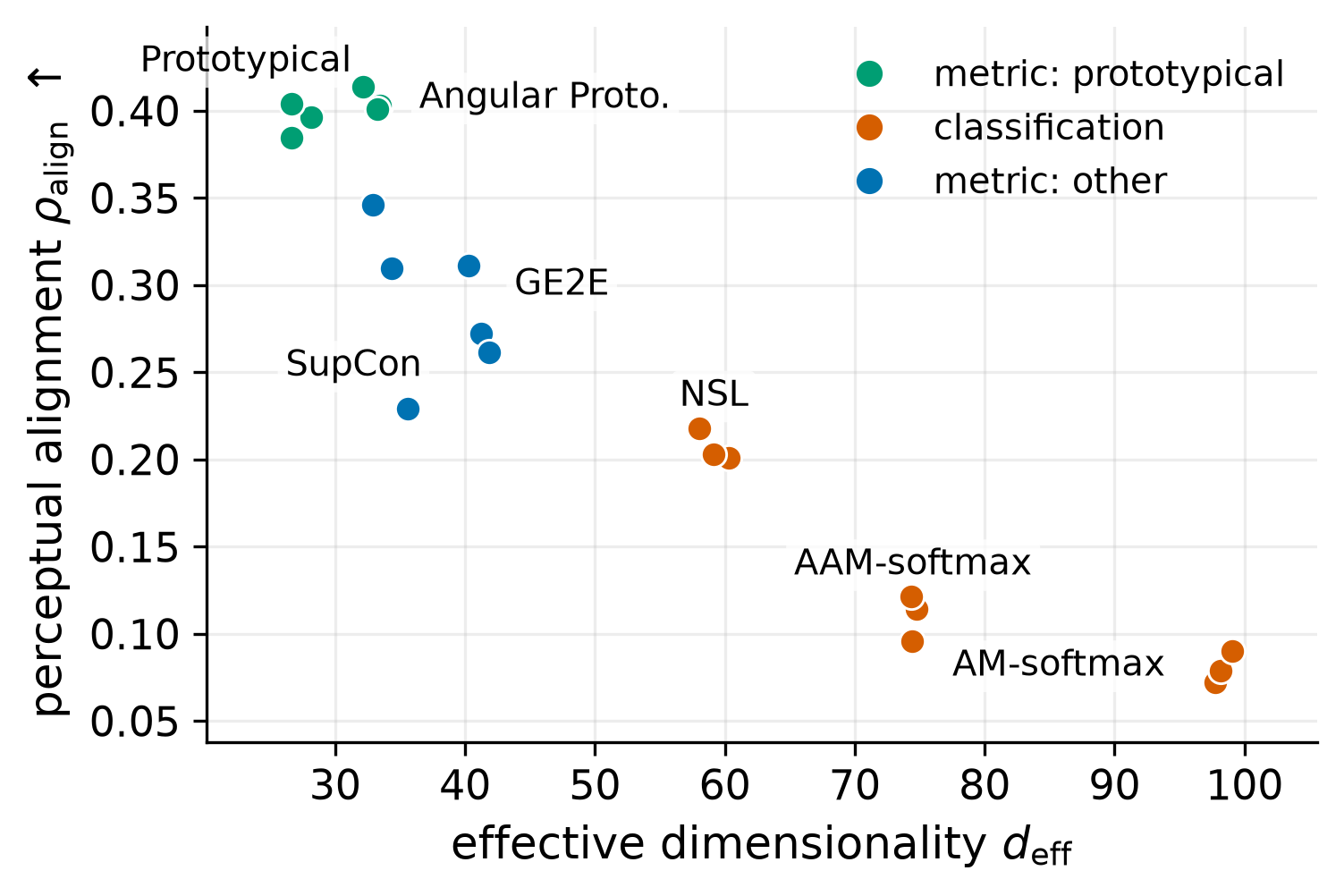}}
\caption{Effective dimensionality ($\effdim$) tracks human perceptual alignment ($\rhoal$). ECAPA, 7 losses $\times$ 3 seeds (21 model points): Spearman$(\effdim, \rhoal) = -0.95$.}
\label{fig:money}
\end{figure}

\subsection{Effective Dimensionality tracks Perceptual Alignment}
\label{ssec:study_e}

Sec.~\ref{ssec:study_a} showed that the training objective decides alignment. As discussed in Sec.~\ref{sec:related_work}, different objectives fundamentally operate by reshaping the spatial distribution and dimensional spread of the learned representations.
We therefore measure this geometry directly using the effective dimensionality \effdim{} (Sec.~\ref{ssec:effdim}) and find that it strongly correlates with the alignment score. Table~\ref{tab:main} shows the \effdim{} on each training combination, and table~\ref{tab:perarch} correlates \effdim{} with \rhoal{} inside each of the five model conditions.
The Spearman coefficient ranges from $-0.71$ to $-1.00$, and the permutation test rejects zero correlation in four of the five model conditions ($p\leq.007$; ReDimNet-B2 $p{=}.088$). Pooled across the 35 training conditions the correlation reaches $-0.95$.
Fig.~\ref{fig:money} isolates the ECAPA-TDNN model condition: its 21 runs fall tightly along the diagonal, and the seven condition means correlate at $-0.96$.
Across all model conditions, models that distribute speakers over fewer directional variations correlate far better with human judgment. This aligns with findings that human judgments of voice similarity can be captured by low-dimensional spaces \cite{baumann2010perceptual}.
Consequently, \effdim{} provides a geometric indicator of perceptual alignment that is computed solely from speaker centroids and requires no human similarity ratings.

\subsection{Dimension Bottlenecks Can Further Improve Perceptual Alignment}
\label{ssec:study_g}

If lower \effdim{} means higher alignment, constraining it during training
should also raise alignment. We test this by re-training ECAPA-TDNN at
$n_{\mathrm{out}}\in\{512,32,10,3\}$ with four objectives
(Sec.~\ref{ssec:conditions}; Table~\ref{tab:bottleneck}). Widening the embedding from the native 192 to 512 changes almost nothing in EER, \rhoal{}, or \effdim{}. The embedding dimension is only an upper bound, and unused directions do not alter the geometry.

Once the dimension is constrained below the intrinsic dimensionality of the
learned representations, \effdim{} is compressed and \rhoal{} rises for all four
objectives. The largest shift is AM-Softmax: the worst-aligned model in the
experiment at 192 dimensions ($0.080$) becomes the best-aligned in the study at
$n_{\mathrm{out}}{=}3$ ($0.738$, $\effdim{=}1.5$)\footnote{Its EER of 20.5\% is comparable to the reported human range, who verify at 17.7\% on VoxSim and at
15.8--26.5\% on VoxCeleb1 \cite{ahn2024voxsim}.}. Restricting capacity also
impairs speaker discrimination, so EER rises at every step, to 13--37\% at
$n_{\mathrm{out}}{=}3$. This trade-off reinforces that a lower EER does not
imply a more human-like representation. At $n_{\mathrm{out}}{=}3$ training also
becomes unstable: AAM-Softmax fails to converge in two of three seeds
($\effdim\approx1.1$, EER $>40$\%), hence its large standard deviations and the
only drop in \rhoal{}. Compressing the embedding dimension during training is
thus an effective geometric lever on perceptual alignment.

\vspace{-1ex}
\begin{table}[htb]
\caption{Embedding dimension bottleneck on ECAPA-TDNN (3-seed mean $\pm$ sd; EER on VoxCeleb1-O; \effdim{} seed sd $\le 2.7$). $\ast$: native width. $^{\dag}$: two of three seeds collapse to $\effdim\approx1.1$ and EER $>40$\%.
\rhoal{} (\sw{blue!41}{blue}) and EER (\sw{red!60}{red}) are shaded as a heat map, darker meaning larger.}
\label{tab:bottleneck}
\centering
\footnotesize
\setlength{\tabcolsep}{0pt}
\begin{tabular*}{\columnwidth}{@{\extracolsep{\fill}}lccc@{\hspace{10pt}}ccc@{}}
\toprule
 & \multicolumn{3}{c}{AAM-Softmax} & \multicolumn{3}{c}{AM-Softmax}\\
\cmidrule(r){2-4}\cmidrule(l){5-7}
$n_{\mathrm{out}}$ & EER\%$\downarrow$ & \rhoal$\uparrow$ & \effdim & EER\%$\downarrow$ & \rhoal$\uparrow$ & \effdim\\
\midrule
512 & \cellcolor{red!2}1.43{\tiny$\pm$.03} & \cellcolor{blue!6}0.107{\tiny$\pm$.007} & 75.8 & \cellcolor{red!2}1.56{\tiny$\pm$.02} & \cellcolor{blue!4}0.079{\tiny$\pm$.014} & 101.6\\
192$^{\ast}$ & \cellcolor{red!2}1.47{\tiny$\pm$.12} & \cellcolor{blue!6}0.111{\tiny$\pm$.013} & 74.5 & \cellcolor{red!2}1.62{\tiny$\pm$.05} & \cellcolor{blue!4}0.080{\tiny$\pm$.009} & 98.3\\
32 & \cellcolor{red!3}2.31{\tiny$\pm$.25} & \cellcolor{blue!16}0.300{\tiny$\pm$.006} & 22.7 & \cellcolor{red!4}2.74{\tiny$\pm$.84} & \cellcolor{blue!14}0.252{\tiny$\pm$.010} & 23.9\\
10 & \cellcolor{red!21}13.99{\tiny$\pm$7.24} & \cellcolor{blue!38}0.685{\tiny$\pm$.111} & 3.1 & \cellcolor{red!10}6.63{\tiny$\pm$.41} & \cellcolor{blue!28}0.501{\tiny$\pm$.029} & 5.4\\
3 & \cellcolor{red!56}37.32{\tiny$\pm$12.93}$^{\dag}$ & \cellcolor{blue!16}0.287{\tiny$\pm$.401} & 1.2 & \cellcolor{red!31}20.48{\tiny$\pm$.35} & \cellcolor{blue!41}0.738{\tiny$\pm$.018} & 1.5\\
\midrule
 & \multicolumn{3}{c}{Angular Prototypical} & \multicolumn{3}{c}{Prototypical}\\
\cmidrule(r){2-4}\cmidrule(l){5-7}
$n_{\mathrm{out}}$ & EER\%$\downarrow$ & \rhoal$\uparrow$ & \effdim & EER\%$\downarrow$ & \rhoal$\uparrow$ & \effdim\\
\midrule
512 & \cellcolor{red!2}1.49{\tiny$\pm$.06} & \cellcolor{blue!22}0.398{\tiny$\pm$.010} & 33.6 & \cellcolor{red!3}2.02{\tiny$\pm$.06} & \cellcolor{blue!21}0.387{\tiny$\pm$.009} & 27.3\\
192$^{\ast}$ & \cellcolor{red!2}1.47{\tiny$\pm$.04} & \cellcolor{blue!22}0.406{\tiny$\pm$.007} & 33.0 & \cellcolor{red!3}1.92{\tiny$\pm$.04} & \cellcolor{blue!22}0.395{\tiny$\pm$.010} & 27.1\\
32 & \cellcolor{red!3}2.15{\tiny$\pm$.09} & \cellcolor{blue!25}0.459{\tiny$\pm$.006} & 16.0 & \cellcolor{red!4}2.61{\tiny$\pm$.10} & \cellcolor{blue!26}0.464{\tiny$\pm$.005} & 13.9\\
10 & \cellcolor{red!8}5.49{\tiny$\pm$.13} & \cellcolor{blue!29}0.529{\tiny$\pm$.018} & 6.6 & \cellcolor{red!10}6.40{\tiny$\pm$.10} & \cellcolor{blue!31}0.564{\tiny$\pm$.007} & 5.6\\
3 & \cellcolor{red!20}13.60{\tiny$\pm$.61} & \cellcolor{blue!37}0.666{\tiny$\pm$.006} & 2.1 & \cellcolor{red!20}13.06{\tiny$\pm$.34} & \cellcolor{blue!36}0.654{\tiny$\pm$.008} & 1.9\\
\bottomrule
\end{tabular*}
\end{table}

\vspace{-2.25ex}
\section{Conclusion}
\label{sec:conclusion}

This study demonstrates that speaker verification performance (EER) is a fundamentally flawed proxy for human perceptual similarity. Instead, perceptual alignment is dictated by the geometry of the embedding space. We identify effective dimensionality ($\effdim$) as a highly reliable indicator of this alignment that requires no human annotation. Representations that distribute speakers across fewer directions naturally mirror the low-dimensional structure of human voice perception. Furthermore, explicitly bottlenecking the embedding dimension during training forcefully compresses $\effdim$ and drastically improves perceptual alignment, albeit at the direct expense of verification accuracy. These findings challenge the community's default reliance on EER and establish $\effdim$ as a principled geometric criterion for evaluating and optimizing speaker embeddings in speech generation tasks.

\vfill\pagebreak


\begin{thebibliography}{10}

\bibitem{deja2022similarity}
K.~Deja, A.~Sanchez, J.~Roth, and M.~Cotescu,
\newblock ``Automatic evaluation of speaker similarity,''
\newblock in {\em Proc. Interspeech}, 2022.

\bibitem{valle1}
C.~Wang, S.~Chen, Y.~Wu, Z.~Zhang, L.~Zhou, S.~Liu, Z.~Chen, Y.~Liu, H.~Wang,
  J.~Li, L.~He, S.~Zhao, and F.~Wei,
\newblock ``Neural codec language models are zero-shot text to speech
  synthesizers,''
\newblock {\em arXiv preprint arXiv:2301.02111}, 2023.

\bibitem{valle2}
S.~Chen, S.~Liu, L.~Zhou, Y.~Liu, X.~Tan, J.~Li, S.~Zhao, Y.~Qian, and F.~Wei,
\newblock ``{VALL-E 2}: Neural codec language models are human parity zero-shot
  text to speech synthesizers,''
\newblock {\em arXiv preprint arXiv:2406.05370}, 2024.

\bibitem{seedtts}
P.~Anastassiou, J.~Chen, J.~Chen, Y.~Chen, Z.~Chen, et~al.,
\newblock ``{Seed-TTS}: A family of high-quality versatile speech generation
  models,''
\newblock {\em arXiv preprint arXiv:2406.02430}, 2024.

\bibitem{cosyvoice}
Z.~Du, Q.~Chen, S.~Zhang, K.~Hu, H.~Lu, Y.~Yang, H.~Hu, S.~Zheng, Y.~Gu, Z.~Ma,
  Z.~Gao, and Z.~Yan,
\newblock ``{CosyVoice}: A scalable multilingual zero-shot text-to-speech
  synthesizer based on supervised semantic tokens,''
\newblock {\em arXiv preprint arXiv:2407.05407}, 2024.

\bibitem{gerlach2020similarity}
L.~Gerlach, K.~McDougall, F.~Kelly, A.~Alexander, and F.~Nolan,
\newblock ``Exploring the relationship between voice similarity estimates by
  listeners and by an automatic speaker recognition system incorporating
  phonetic features,''
\newblock {\em Speech Communication}, vol. 124, pp. 85--95, 2020.

\bibitem{liu2024comparison}
S.~Liu, M.~Babel, and J.~Zhu,
\newblock ``A comparison of voice similarity through acoustics, human
  perception and deep neural network ({DNN}) speaker verification systems,''
\newblock in {\em Proc. Interspeech}, 2024, pp. 3674--3678.

\bibitem{das2020predictions}
Rohan~Kumar Das, Tomi Kinnunen, Wen-Chin Huang, Zhenhua Ling, Junichi
  Yamagishi, Yi~Zhao, Xiaohai Tian, and Tomoki Toda,
\newblock ``Predictions of subjective ratings and spoofing assessments of voice
  conversion challenge 2020 submissions,''
\newblock in {\em Proceedings of the Joint Workshop for the Blizzard Challenge
  and Voice Conversion Challenge 2020}, 2020.

\bibitem{chung2020defence}
J.~S. Chung, J.~Huh, S.~Mun, M.~Lee, H.~S. Heo, S.~Choe, C.~Ham, S.~Jung, B.-J.
  Lee, and I.~Han,
\newblock ``In defence of metric learning for speaker recognition,''
\newblock in {\em Proc. Interspeech}, 2020.

\bibitem{i-vector}
Najim Dehak, Patrick~J. Kenny, Réda Dehak, Pierre Dumouchel, and Pierre
  Ouellet,
\newblock ``Front-end factor analysis for speaker verification,''
\newblock {\em IEEE Transactions on Audio, Speech, and Language Processing},
  vol. 19, no. 4, pp. 788--798, 2011.

\bibitem{snyder2018xvector}
D.~Snyder, D.~Garcia-Romero, G.~Sell, D.~Povey, and S.~Khudanpur,
\newblock ``X-vectors: Robust {DNN} embeddings for speaker recognition,''
\newblock in {\em Proc. IEEE International Conference on Acoustics, Speech and
  Signal Processing (ICASSP)}, 2018.

\bibitem{wang2018amsoftmax}
F.~Wang, W.~Liu, H.~Liu, and J.~Cheng,
\newblock ``Additive margin softmax for face verification,''
\newblock {\em IEEE Signal Processing Letters}, vol. 25, no. 7, pp. 926--930,
  2018.

\bibitem{deng2019arcface}
J.~Deng, J.~Guo, N.~Xue, and S.~Zafeiriou,
\newblock ``{ArcFace}: Additive angular margin loss for deep face
  recognition,''
\newblock in {\em Proc. IEEE/CVF Conference on Computer Vision and Pattern
  Recognition (CVPR)}, 2019.

\bibitem{wan2018ge2e}
L.~Wan, Q.~Wang, A.~Papir, and I.~Lopez Moreno,
\newblock ``Generalized end-to-end loss for speaker verification,''
\newblock in {\em Proc. IEEE International Conference on Acoustics, Speech and
  Signal Processing (ICASSP)}, 2018.

\bibitem{snell2017proto}
J.~Snell, K.~Swersky, and R.~Zemel,
\newblock ``Prototypical networks for few-shot learning,''
\newblock in {\em Advances in Neural Information Processing Systems}, 2017,
  vol.~30.

\bibitem{khosla2020supcon}
P.~Khosla, P.~Teterwak, C.~Wang, A.~Sarna, Y.~Tian, P.~Isola, A.~Maschinot,
  C.~Liu, and D.~Krishnan,
\newblock ``Supervised contrastive learning,''
\newblock in {\em Advances in Neural Information Processing Systems}, 2020,
  vol.~33.

\bibitem{jing2022dimcollapse}
L.~Jing, P.~Vincent, Y.~LeCun, and Y.~Tian,
\newblock ``Understanding dimensional collapse in contrastive self-supervised
  learning,''
\newblock in {\em Proc. International Conference on Learning Representations
  (ICLR)}, 2022.

\bibitem{roth2020revisiting}
K.~Roth, T.~Milbich, S.~Sinha, P.~Gupta, B.~Ommer, and J.~P. Cohen,
\newblock ``Revisiting training strategies and generalization performance in
  deep metric learning,''
\newblock in {\em Proc. International Conference on Machine Learning (ICML)},
  2020.

\bibitem{baumann2010perceptual}
Oliver Baumann and Pascal Belin,
\newblock ``Perceptual scaling of voice identity: common dimensions for
  different vowels and speakers,''
\newblock {\em Psychological Research}, vol. 74, no. 1, pp. 110--120, 2010.

\bibitem{gao2017theory}
P.~Gao, E.~Trautmann, B.~Yu, G.~Santhanam, S.~Ryu, K.~Shenoy, and S.~Ganguli,
\newblock ``A theory of multineuronal dimensionality, dynamics and
  measurement,''
\newblock {\em bioRxiv preprint 214262}, 2017,
\newblock doi:10.1101/214262.

\bibitem{delgiudice2021ed}
M.~{Del Giudice},
\newblock ``Effective dimensionality: A tutorial,''
\newblock {\em Multivariate Behavioral Research}, vol. 56, no. 3, pp. 527--542,
  2021.

\bibitem{desplanques2020ecapa}
B.~Desplanques, J.~Thienpondt, and K.~Demuynck,
\newblock ``{ECAPA-TDNN}: Emphasized channel attention, propagation and
  aggregation in {TDNN} based speaker verification,''
\newblock in {\em Proc. Interspeech}, 2020.

\bibitem{yakovlev2024redimnet}
I.~Yakovlev, R.~Makarov, A.~Balykin, P.~Malov, A.~Okhotnikov, and N.~Torgashov,
\newblock ``Reshape dimensions network for speaker recognition,''
\newblock in {\em Proc. Interspeech}, 2024.

\bibitem{chen2022wavlm}
S.~Chen, C.~Wang, Z.~Chen, Y.~Wu, S.~Liu, Z.~Chen, J.~Li, N.~Kanda,
  T.~Yoshioka, X.~Xiao, et~al.,
\newblock ``{WavLM}: Large-scale self-supervised pre-training for full stack
  speech processing,''
\newblock {\em IEEE Journal of Selected Topics in Signal Processing}, vol. 16,
  no. 6, pp. 1505--1518, 2022.

\bibitem{chung2018voxceleb2}
J.~S. Chung, A.~Nagrani, and A.~Zisserman,
\newblock ``{VoxCeleb2}: Deep speaker recognition,''
\newblock in {\em Proc. Interspeech}, 2018.

\bibitem{wang2018cosface}
Hao Wang, Yitong Wang, Zheng Zhou, Xing Ji, Dihong Gong, Jingchao Zhou, Zhifeng
  Li, and Wei Liu,
\newblock ``{CosFace}: Large margin cosine loss for deep face recognition,''
\newblock in {\em Proc. IEEE/CVF CVPR}, 2018, pp. 5265--5274.

\bibitem{nagrani2017voxceleb}
A.~Nagrani, J.~S. Chung, and A.~Zisserman,
\newblock ``{VoxCeleb}: A large-scale speaker identification dataset,''
\newblock in {\em Proc. Interspeech}, 2017.

\bibitem{nagrani2020voxceleb}
A.~Nagrani, J.~S. Chung, W.~Xie, and A.~Zisserman,
\newblock ``Voxceleb: Large-scale speaker verification in the wild,''
\newblock {\em Computer Speech \& Language}, vol. 60, pp. 101027, 2020.

\bibitem{ahn2024voxsim}
J.~Ahn, Y.~Kim, Y.~Choi, D.~Kwak, J.-H. Kim, S.~Mun, and J.~S. Chung,
\newblock ``{VoxSim}: A perceptual voice similarity dataset,''
\newblock in {\em Proc. Interspeech}, 2024.

\bibitem{ravanelli2021speechbrain}
M.~Ravanelli, T.~Parcollet, P.~Plantinga, A.~Rouhe, S.~Cornell, et~al.,
\newblock ``{SpeechBrain}: A general-purpose speech toolkit,''
\newblock {\em arXiv preprint arXiv:2106.04624}, 2021.

\end{thebibliography}
\end{document}